\documentclass[prd,aps,showpacs,nofootinbib,preprint,eqsecnum]{revtex4}
\usepackage{graphicx,color,amsmath,amsxtra}
\usepackage{amssymb}
\usepackage[english]{babel}
\usepackage{amsfonts}

\allowdisplaybreaks[4]
\begin{document}

\title{Cosmological perturbations in mimetic matter model}

\author{Jiro Matsumoto$^{1}$\footnote{E-mail address: jmatsumoto@kpfu.ru},
Sergei D.~Odintsov$^{2,3}$, and Sergey V.~Sushkov$^{1}$}
\affiliation{
$^1$Institute of Physics, Kazan Federal University, Kremlevskaya Street 18,
Kazan 420008, Russia\\
$^2$Consejo Superior de Investigaciones Cient\'{\i}ficas, ICE/CSIC-IEEC,
Campus UAB, Facultat de Ci\`{e}ncies, Torre C5-Parell-2a pl, E-08193
Bellaterra (Barcelona), Spain\\
$^3$Instituci\'{o} Catalana de Recerca i Estudis Avan\c{c}ats
(ICREA), Barcelona, Spain\\
}

\begin{abstract}
We investigate the cosmological evolution of mimetic matter model with 
arbitrary scalar potential. The cosmological reconstruction
, which is the way to construct a model for arbitrary evolutions of the scale factor, 
is explicitly 
done for different choices of potential. 
The cases that mimetic matter 
model shows the evolution as Cold Dark 
Matter(CDM), wCDM model, dark matter and dark energy with dynamical 
$Om(z)$
, where $Om(z) \equiv [ ( H(z)/H_0 )^2-1 ]/[(1+z)^3 -1],$
 or phantom dark energy with phantom-non-phantom crossing are 
presented in detail. The cosmological perturbations for such evolution are 
studied in mimetic matter model.
For instance, the evolution behavior of the matter density contrast which 
is different from usual
one, i.e.
$\ddot \delta + 2 H \dot \delta - \kappa ^2 \rho \delta /2 = 0$ is 
investigated. 
The possibility of peculiar evolution of $\delta$ in the model under 
consideration is shown. Special attention is paid to the behavior of 
matter density contrast near to future singularity where decay of 
perturbations may occur much earlier the singularity.

\end{abstract}


\maketitle
\section{Introduction}

The existence of dark energy and dark matter is indirectly shown by
   the observations of Cosmic Microwave Background (CMB),
Large Scale Structure (LSS) of the Universe, Super Novae of Type Ia (SNIa), and
related observational probes.
Dark energy is usually regarded as a cosmological constant $\Lambda$ in the
standard model of cosmology, $\Lambda$CDM model.
However, in general it represents some energy which have 
a negative pressure with equation of state parameter 
being less than $-1/3$. 
On the other hand, dark matter, in particular Cold Dark Matter (CDM) means the
nonrelativistic unknown matter.
There are many candidates for dark energy (for review,
see\cite{Bamba:2012cp,review}) and dark
matter\cite{Bertone:2004pz,Feng:2010gw}.
However,
it is not yet clear which model is viable one.

In this paper, we will consider mimetic matter model
\cite{Chamseddine:2013kea},
which was first proposed as a model of dark matter. However, it was realized
later that it can be treated
as a model of dark energy by adding the potential term of the scalar field
\cite{Chamseddine:2014vna}.
Mimetic matter model is a conformally invariant theory by making the physical
metric as a
product of an auxiliary metric and the contraction of an auxiliary metric and
the kinetic term of the scalar field.
Whereas, the equivalence between mimetic matter model and
a scalar field model with a Lagrange multiplier \cite{DDE} is shown in
\cite{Golovnev:2013jxa,Barvinsky:2013mea}.
Therefore, we treat mimetic matter model as a scalar field model with a
Lagrange multiplier in this paper.

The contents of the paper are the following.
In Sec.~II, we  consider the background evolution of the Universe in general
mimetic matter model. The cosmological reconstruction of the model 
, which is the way to construct the action which may realize the given 
evolution history of the universe by the choice of potential, 
is done and
$\Lambda$CDM-like era is explicitly executed for corresponding scalar potential.
The evolution of the matter density perturbation is discussed in Sec.~III.
First, mimetic matter model without potential is used to evaluate whether
mimetic dark matter
also behaves as dark matter at the cosmological perturbations level.
Next, we investigate the matter density perturbation in general mimetic matter
model with scalar potential.
In Sec.~IV, the possibility of the phantom -- non-phantom transition is
explicitly demonstrated in mimetic matter model.
The behavior of the cosmological perturbations is also discussed, especially
the behavior close to finite-time future singularity.
Concluding remarks are given in Sec.~V.
The units of $k_\mathrm{B} = c = \hbar = 1$ are used and
gravitational constant $8 \pi G$ is denoted by
${\kappa}^2 \equiv 8\pi/{M_{\mathrm{Pl}}}^2$
with the Planck mass of $M_{\mathrm{Pl}} = G^{-1/2} = 1.2 \times 10^{19}$GeV in
this paper.
\section{The accelerating evolution of the isotropic and homogeneous universe}
We  consider the action of  mimetic matter model\cite{Chamseddine:2014vna}:
\begin{align}
S= \int d^4 x \sqrt{-g} \left [ \frac{1}{2 \kappa ^2} R -V(\phi) + \lambda
(g^{\mu \nu} \partial _\mu \phi \partial _\nu \phi +1)
   \right ]+S_\mathrm{matter},
\label{10}
\end{align}
where $V$ is an arbitrary function of the scalar field $\phi$
and $\lambda$ is a Lagrange multiplier. The Einstein equation obtained from
Eq.~(\ref{10}) is
\begin{align}
R_{\mu \nu} - \frac{1}{2}g_{\mu \nu} R = - \kappa ^2 g_{\mu \nu}V(\phi)
-2 \kappa ^2 \lambda \partial _\mu \phi \partial _\nu \phi + \kappa ^2 g_{\mu
\nu} \lambda (g^{\rho \sigma} \partial _\rho \phi \partial _\sigma \phi +1)
+ \kappa ^2 T_{\mu \nu},
\label{EE}
\end{align}
where $R_{\mu \nu} = \partial _\sigma \Gamma ^\sigma _{\mu \nu} - \partial _\mu
\Gamma ^\sigma _{\nu \sigma}
+ \Gamma ^\sigma _{\mu \nu} \Gamma ^\rho _{ \sigma \rho}
- \Gamma ^\sigma _{\mu \rho} \Gamma ^\rho _{\nu \sigma} $
and $T_{\mu \nu}$ is the energy momentum tensor of the usual matter.
On the other hand, the equation given by the variation with respect to the
scalar field $\phi$ is the following one:
\begin{equation}
-V_{, \phi} -2 \nabla ^\mu (\lambda \partial _\mu \phi)=0,
\label{FE}
\end{equation}
where $V_{, \phi}$ $\equiv$ $dV(\phi)/d \phi$. The constraint equation given by
the variation of $\lambda$ is
\begin{equation}
g^{\mu \nu} \partial _\mu \phi \partial _\nu \phi +1 =0.
\label{LAMBDA}
\end{equation}
When  the spatially flat Friedmann-Lemaitre-Robertson-Walker (FLRW) metric,
$ds^2 =-dt^2+a^2(t)\sum_{i=1}^3 dx^i dx^i)$, is taken,
FLRW equations are written by
\begin{align}
3H^2 =  \kappa ^2 \rho -2 \kappa ^2 \lambda \dot \phi ^2 +\kappa ^2 V + \kappa
^2 \lambda (\dot \phi ^2 -1)\label{FL00}, \\
-2 \dot H -3H^2 =   \kappa ^2 w \rho - \kappa ^2 V - \kappa ^2 \lambda (\dot
\phi ^2 -1)
\label{FLii},
\end{align}
where $H \equiv \dot a(t) / a(t)$.
$\rho$ is the energy-density of the matter and $w$ is the equation of state
(EoS) parameter expressed by $w=p/ \rho$.
From Eqs.~(\ref{FE}) and (\ref{LAMBDA}) we have
\begin{align}
2 \dot \lambda \dot \phi + 2 \lambda \ddot \phi +6H \lambda \dot \phi - V_{,
\phi} =0, \label{BFE}\\
\dot \phi ^2 = 1. \label{BLA}
\end{align}
The equation of continuity of the matter is
\begin{equation}
\dot \rho +3(1+w) H \rho = 0.
\label{15}
\end{equation}
If $w$ is constant, one can integrate Eq.~(\ref{15}) as
\begin{equation}
\rho = \rho _0 \left ( \frac{a}{a_0} \right ) ^{-3(1+w)},
\label{sol1}
\end{equation}
where $\rho _0 a_0^{3(1+w)}$ is an integration constant.
While, combining Eqs.~(\ref{FL00}), (\ref{FLii}), and (\ref{BLA}) it gives the
equations
\begin{equation}
\lambda = \frac{1}{\kappa ^2}\dot H + \frac{1}{2}(1+w) \rho
\label{sol2}
\end{equation}
and
\begin{equation}
V=  w \rho + \frac{1}{\kappa ^2} (2 \dot H + 3H^2).
\label{sol3}
\end{equation}
Therefore, all the solutions for the variables will be found if
Eq.~(\ref{sol3}) can be solved with respect to $a(t)$.
From the other side, one can say that this model is able to realize arbitrary
evolution history of the Universe by
using the arbitrariness of the function $V(\phi)$.
Indeed, one can choose the form of $V(\phi)$ for the arbitrary function $f$
such that $a(t)=f(t)$ as
\begin{equation}
V(\phi) = w \rho _0 \left ( \frac{f(\phi)}{f(\phi _0)} \right ) ^{-3(1+w)}
+ \frac{1}{\kappa ^2} \left ( 2 \frac{f''(\phi)}{f(\phi)} +
\frac{f'(\phi)^2}{f(\phi)^2} \right ).
\label{rec}
\end{equation}

\subsection{Cosmological reconstruction}
Some forms of the function $V(\phi)$ are considered in
ref.\cite{Chamseddine:2014vna}.
Here, we  consider some other examples of $V(\phi)$, which are adequate to the
cosmological
observations and yield the accelerating universe. In particularly, the form of
$V(\phi)$ which can yield the phantom -- non-phantom transition is
treated in Sec.~\ref{phantom}. In the following, we assume the value of EoS
parameter of the usual matter as
$p_\mathrm{matter}/ \rho _\mathrm{matter} = w_\mathrm{matter} = 0$. 

\subsubsection{Mimetic wCDM model}
wCDM model consists of cold dark matter and a fluid with EoS parameter $w<0$.
For the case $w=-1$, wCDM model is equivalent to the $\Lambda$CDM model,
so its background evolution  can be realized by choosing the form of the
potential as $V=\Lambda$.
In the case of $w \neq -1$, we can reproduce the background evolution of the
Universe in wCDM model
if the effective energy density $-2 \lambda + V$ and the effective pressure
$-V$ satisfy the following conditions:
\begin{equation}
-2 \lambda + V = \rho _{dm0} a^{-3} + \rho _{w0} a^{-3(1+w)}  \label{wenergy}
\end{equation}
and
\begin{equation}
-V = w \rho _{w0} a^{-3(1+w)}, \label{wpressure}
\end{equation}
where $\rho _{dm0}$ and $\rho _{w0}$ are the current energy-density of dark
matter and that of a fluid, respectively.
Here,  the current value of the
scale factor is chosen as $a_0=1$.
Thus,
\begin{align}
\lambda = - \frac{1}{2} \left \{ \rho _{dm0} a^{-3} + (1+w) \rho _{w0}
a^{-3(1+w)} \right \}, \label{wlambda} \\
V = -w \rho _{w0} a^{-3(1+w)}. \label{wpotential}
\end{align}
The specific form of the potential is very complicated even if it can be
obtained by solving Eq.~(\ref{FL00}).
However, one can easily obtain some approximated forms of $V(\phi)$.

\begin{figure}
\begin{center}
\includegraphics[clip, width=0.8\columnwidth]{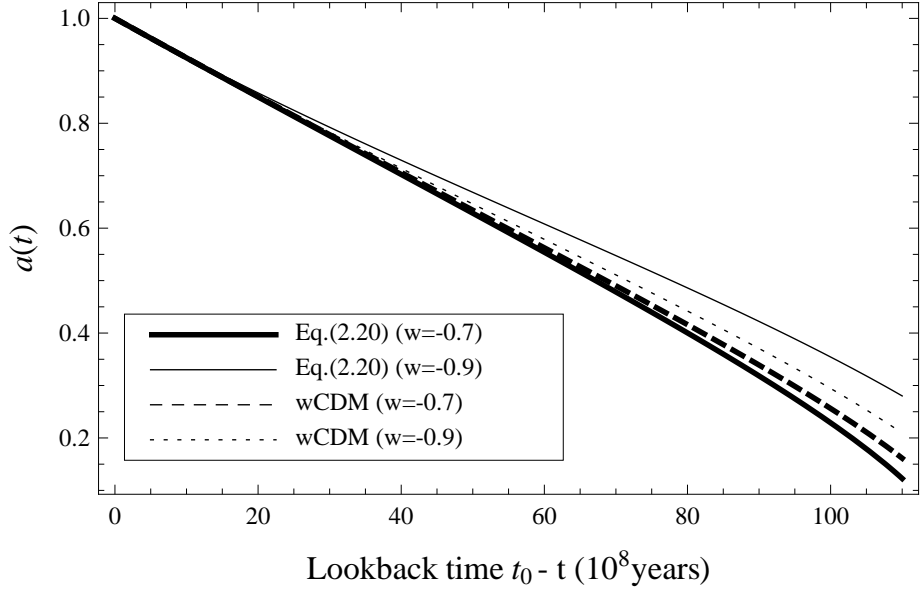}
\end{center}
\caption{Comparison of the time dependence of the scale factor between wCDM model and Eq.~(\ref{imdapp}). 
Thick solid line, solid line, dashed line, and dotted line represent Eq.~(\ref{imdapp}) with $w=-0.7$, 
Eq.~(\ref{imdapp}) with $w=-0.9$, wCDM model with $w=-0.7$, and wCDM model with $w=-0.9$, respectively. 
$H_0 = 74$ km/s/Mpc, $\Omega _{m0} = 0.3$, and $\Omega _{w0} =0.7$ are assumed. }
\label{app1}
\end{figure}

When $w>-1$, an approximated solution of wCDM model is given by
\begin{equation}
a(t)= \left [ \frac{3}{4}\kappa ^2 (\rho _{b0} + \rho _{dm0}) \right ] ^{\frac{1}{3}}
t^{\frac{2}{3}}
\left \{ 1+ \left [ (1+w)^2 \frac{\rho _{w0}}{\rho _{b0} + \rho _{dm0}} \left (
\frac{3}{4}\kappa ^2 (\rho _{b0} + \rho _{dm0}) t^2 \right )^{-w} \right ]^\frac{1}{3(1+w)}
\right \}
, \label{appa}
\end{equation}
where $\rho _{b0}$ is the current energy-density of the baryonic matter, 
because the solutions of Eq.~(\ref{FL00}) in the matter dominant universe and
in the fluid matter dominant universe are given by 
$a=(3\kappa ^2 (\rho _{b0} + \rho _{dm0})/4)^{1/3} t^{2/3}$ 
and $a=[3(1+w)^2 \kappa ^2 \rho _{w0}/4]^{1/(3(1+w))}t^{2/(3(1+w))}$, respectively.
Therefore, an approximated form of potential $V(\phi)$ is expressed as 
\begin{align}
V(\phi) = \frac{4}{9 \kappa ^2 \phi ^2} \frac{1}
{1+ \left [ (1+w)^2 \frac{\rho _{w0}}{\rho _{b0} + \rho _{dm0}} \left (  \frac{3}{4}\kappa ^2
(\rho _{b0} + \rho _{dm0}) \phi ^2 \right )^{-w} \right ]^{-\frac{1}{3(1+w)}}}
\nonumber \\
\times \left \{ - \frac{3w + w^2}{(1+w)^2} + \frac{w^2}{(1+w)^2}
   \frac{1}
{1+ \left [ (1+w)^2 \frac{\rho _{w0}}{\rho _{b0} + \rho _{dm0}} \left (  \frac{3}{4}\kappa ^2
(\rho _{b0} + \rho _{dm0}) \phi ^2 \right )^{-w} \right ]^{-\frac{1}{3(1+w)}}}
\right \} . \label{appV}
\end{align}
Eq.~(\ref{appa}) is a very rough approximation as a kind of the interpolation. 
To improve the accuracy of the approximation, one can use the current value of the Hubble rate $H_0$ as: 
\begin{align}
a(t)=\left [ \frac{3}{4}\kappa ^2 (\rho _{b0} + \rho _{dm0}) t^2 \right ]^{\frac{1}{3}}
+ \left [ \frac{3}{4}(1+w)^2\kappa ^2 \rho _{w0} t^2 \right ]^{\frac{1}{3(1+w)}} 
+ C (H_0 t)^{\frac{2+w}{3(1+w)}}, 
\label{imdapp}
\end{align}
where $C$ is a constant tuned to satisfy $\dot a / a = H_0$ when $a(t)=1$. 
The value of $C$ is evaluated by solving a simultaneous equation $\{ a(t_0)=1$, $H(t_0)=H_0 \}$ with respect to $t_0$ and $C$.  
In the case of $w=-0.7$, $t_0=116.752 \times 10^8$ years and $C=0.126265$ are obtained. 
The comparisons between the exact wCDM model and Eq.~(\ref{imdapp}) when $w=-0.7$ and $w=-0.9$ are shown in Fig.~\ref{app1}. 
It seems that Eq.~(\ref{imdapp}) is not a bad approximation when $w$ is large, 
although only three points $t=+0,t_0, + \infty$ are used as the interpolation points. 
\subsubsection{Reconstruction with respect to $Om$}
One can rewrite Eqs.~(\ref{sol2}) and (\ref{sol3}) by using $Om(z)$ and
$Om'(z)$ instead of
$H(t)$ and $H'(t)$ as
\begin{align}
\lambda = \frac{H_0^2}{\kappa ^2} \left \{ -3/2 (1+z)^3 Om - \frac{1}{2}(1+z)^4
Om' + \frac{1}{2} (1+z) Om' \right \}
+ \frac{1}{2} \rho _{b0} (1+z)^3, \label{om1} \\
V=\frac{H_0^2}{\kappa ^2} \left \{ 3(1-Om) -(1+z)^4 Om' + (1+z)Om' \right \} ,
\label{om2}
\end{align}
where $\rho _{b0}$ is  current energy-density of the baryons. $Om(z)$ is
defined by \cite{Sahni:2008xx,Zunckel:2008ti}
\begin{equation}
Om(z) = \frac{\left ( H(z)/H_0 \right )^2-1}{(1+z)^3 -1}. \label{om3}
\end{equation}
In the case of the $\Lambda$CDM model, $Om(z)$ is in accord with the current
matter fraction $\Omega _{m0}$.
However, in the recent paper\cite{Sahni:2014ooa}, a discrepancy between
CMB observations (Planck+WP) and BAO observations is pointed out. It is pointed
out  that the value of $Om h^2$ is larger for larger redshift.
The explicit values of $\Omega _{m0}h^2$ for Planck+WP and $Omh^2$ for BAO
observations are
$\Omega _{m0}h^2 = 0.1426 \pm 0.0025$\cite{Ade:2013zuv} and $Omh^2 \approx
0.122 \pm 0.01$\cite{Sahni:2014ooa}, respectively.
Of course, these values are incompatible in the $\Lambda$CDM model, because
$Omh^2$ is constant with respect to redshift.
However, we can construct a model that satisfies these conditions.
If we consider the following scale factor as a solution of Eqs.~(\ref{FL00})
and (\ref{FLii}):
\begin{align}
a(t)=\left ( \frac{g(t)}{1-g(t)} \right )^{1/3} \sinh ^{2/3} \left [
\frac{3}{2} \sqrt{1-g(t)} H_0 t \right ], \nonumber \\
g(t)=\left ( 0.122 + 0.04 \frac{t_{CMB}}{t} - 0.02 \frac{t_{CMB}^2}{t^2} \right
)/h^2, \label{om4}
\end{align}
where $t_{CMB}$ is a time of $z=1089$, then $Omh^2 \approx 0.122$ for small
redshift and $Omh^2 = 0.142$ at $z=1089$ is realized.
The potential which causes such an expansion history is given by using
Eq.~(\ref{rec}), however, it is not written down explicitly, being very
complicated.
Similarly, we can reconstruct any requested universe history. In the next
section,
the evolution of the matter density perturbations is studied.

\section{Cosmological perturbations with account of  matter \label{per}}
In this section, we investigate the
growth rate of the matter density perturbations (for general review of
cosmological perturbations theory, see \cite{m,g,sasaki}).
When we use the metric of the Newtonian gauge,
$ds^2 = (-1+2\Phi) dt^2+ (1+2\Psi) a^2(t) \sum_{i=1}^3 dx^i dx^i $,
then $(00)$, $(0i)=(i0)$, $(ij)$, $i\neq j$, and $(ii)$ elements of the
linearized Einstein equation in
the Fourier space
are represented as follows, respectively:
\begin{align}
-6H^2 \Phi -2\frac{k^2}{a^2} \Psi -6H \partial _0 \Psi = - \kappa ^2 \delta
\rho + 2\kappa ^2 \delta \lambda
+2 \kappa ^2 \lambda \Phi +2 \kappa ^2 \lambda \delta \dot \phi - \kappa ^2
V_{, \phi} \delta \phi
\label{R00}, \\
2 (H \Phi + \partial _0 \Psi) = \kappa ^2 (\rho + p)\delta u + 2 \kappa ^2
\lambda \delta \phi
\label{R0i}, \\
\Phi - \Psi = 0
\label{Rij}, \\
\left ( \frac{k^2}{a^2} + \frac{\partial _i \partial _i}{a^2} - 2H \partial _0
-4 \dot H -6 H^2 \right ) \Phi
- \left ( \frac{k^2}{a^2} + \frac{\partial _i \partial _i}{a^2} + 2 \partial _0
\partial _0 +6 H \partial _0 \right ) \Psi = \nonumber \\
\kappa ^2 c_\mathrm{s}^2 \delta \rho -2 \kappa ^2 \lambda (\Phi + \delta \dot
\phi) - \kappa ^2 V_{, \phi} \delta \phi
\label{Rii},
\end{align}
where we defined the sound speed by $c_\mathrm{s}^2 \equiv \delta p / \delta
\rho$, the matter density perturbation by
$\delta \equiv \delta \rho / \rho$ and the scalar perturbation of the four
velocity as $\partial _i \delta u$.
The linearized equations of (\ref{FE}) and (\ref{LAMBDA}) are given as follows:
\begin{align}
\delta \dot \lambda + 3H \delta \lambda + \lambda \delta \ddot \phi + (\dot
\lambda + 3H \lambda) \delta \dot \phi
+ \left ( \lambda \frac{k^2}{a^2}  -\frac{1}{2} V_{, \phi \phi} \right ) \delta
\phi \nonumber \\
+ \lambda \dot \Phi +2( \dot \lambda +3H \lambda) \Phi +3 \lambda \dot \Psi =0,
\label{PFE}\\
\dot \phi \delta \dot \phi  + \dot \phi ^2 \Phi = 0 \label{PLA}.
\end{align}
From the perturbation of the equation of continuity, $\nabla _\mu T^{\mu
\nu}=0$, one gets
\begin{align}
\delta \dot \rho + 3H(\delta \rho + \delta p) +a^{-2} \partial _i \{ (\rho +p)
\partial _i \delta u  \}
+ 3 \dot \Psi (\rho + p) = 0\, , \label{T0} \\
a^{-3} \partial _0 \{ a^3 (\rho + p) \partial _i \delta u \} + c_\mathrm{s}^2
\partial _i \delta \rho - (\rho + p) \partial _i \Phi
=0\, . \label{Ti}
\end{align}
In the following, we treat the equation of state parameter and the sound speed
of the matter as $w=c_\mathrm{s}=0$ by assuming
the era after matter dominance.
\subsection{Mimetic dark matter \label{mdm}}
First, we  consider the case $V'(\phi) =0$ as a simple example.
Then, the field $\lambda$ behaves as dark matter in the flat FLRW
space-time\cite{Chamseddine:2013kea}.
Let us investigate whether the field $\lambda$ also behaves as dark matter when
we consider cosmological perturbations.
First, Eqs.~(\ref{PLA}) and (\ref{Rij}) give the relation $\Phi = \Psi =
-\delta \dot \phi$.
By applying the subhorizon approximation, $\partial _0 \sim H$ and $H^2 \ll
k^2/a^2$,
Eqs.~(\ref{R00}) and (\ref{PFE}), then, yield the following equations:
\begin{align}
   2\frac{k^2}{a^2} \delta \dot \phi \simeq - \kappa ^2 \delta \rho + 2\kappa ^2
\delta \lambda
\label{R00sr}, \\
\delta \dot \lambda + 3H \delta \lambda +  \lambda \frac{k^2}{a^2} \delta \phi
\simeq 0, \label{LFEsr}
\end{align}
Differentiating Eq.~(\ref{LFEsr}) with respect to $t$ gives
\begin{equation}
\ddot \delta _\lambda + 2H \dot \delta _\lambda + \frac{k^2}{a^2}\delta \dot
\phi \simeq 0,
\label{LM}
\end{equation}
where $\delta _\lambda \equiv \delta \lambda / \lambda$.
In the case of $w=c_s^2=0$, we can obtain the following equation by using
Eqs.~(\ref{T0}), (\ref{Ti}),
\begin{equation}
\ddot \delta + 2H \dot \delta  + \frac{k^2}{a^2}\delta \dot \phi \simeq 0.
\label{BM}
\end{equation}
Taking into account $\rho, \lambda \propto a^{-3}$,
we obtain
\begin{equation}
\ddot \delta _\mathrm{tot} + 2H \dot \delta _\mathrm{tot} +
\frac{k^2}{a^2}\delta \dot \phi \simeq 0,
\label{tots}
\end{equation}
where $\rho _\mathrm{tot} \equiv \rho - 2 \lambda$ and $\delta _\mathrm{tot}
\equiv (\delta \rho - 2 \delta \lambda)/(\rho - 2 \lambda)$.
By using Eq.~(\ref{R00sr}), we finally obtain
\begin{equation}
\ddot \delta _\mathrm{tot} + 2H \dot \delta _\mathrm{tot}
- \frac{\kappa ^2}{2} \rho _\mathrm{tot} \delta _\mathrm{tot} \simeq 0
\label{totsr}.
\end{equation}
Now, one can recognize the equivalence between mimetic dark matter and fluid
dark matter
at the linear perturbation level because Eq.~(\ref{totsr}) is same as that of
the $\Lambda$CDM model. 
The equivalence of mimetic dark matter and fluid dark matter had 
been shown in Ref.~\cite{DDE}. 
\subsection{Mimetic matter with potential \label{mmp}}
In this section, we consider more involved case $V_{,\phi} \neq 0$.
The perturbation equations (\ref{R0i}), (\ref{Rii}), and (\ref{T0}) can be
simplified as follows by using
Eqs.~(\ref{Rii}) and (\ref{PLA}):
\begin{align}
-2 \delta \ddot \phi -2 H \delta \dot \phi = \kappa ^2 \rho \delta u + 2 \kappa
^2 \lambda \delta \phi
\label{410}, \\
2 \delta \dddot \phi + 8 H \delta \ddot \phi + (4 \dot H +6 H^2 )\delta \dot
\phi =
   - \kappa ^2 V_{, \phi} \delta \phi
\label{420}, \\
\delta \dot \rho + 3H \delta \rho - \frac{k^2}{a^2}\rho \delta u
- 3 \rho \delta \ddot \phi = 0\,
\label{430}.
\end{align}
Eliminating the term proportional to $\delta u$ from Eqs.~(\ref{410}) and
(\ref{430}) gives
\begin{equation}
\dot \delta + \left ( \frac{2}{\kappa ^2 \rho} \frac{k^2}{a^2}-3 \right )
\delta \ddot \phi
+ \frac{2H}{\kappa ^2 \rho} \frac{k^2}{a^2} \delta \dot \phi + \frac{2
\lambda}{\rho} \frac{k^2}{a^2} \delta \phi = 0.
\label{440}
\end{equation}
Then, we can derive the following evolution equation of the matter density
perturbation:
\begin{align}
\ddddot \delta + \left ( 7H - \frac{\dot \lambda}{\lambda} + O \left (
\frac{a^2H^2}{k^2} \right ) \right ) \dddot \delta +
\left ( 16H^2 - 4 \kappa ^2 (\rho - 2 \lambda) -5H \frac{\dot \lambda}{\lambda}
+ O \left ( \frac{a^2H^2}{k^2} \right ) \right ) \ddot \delta \nonumber \\
+ \frac{3}{2} \left ( 8H^3 -3 \kappa ^2 H (\rho - 4 \lambda) -4 H^2 \frac{\dot
\lambda}{\lambda} + \kappa ^2 \rho \frac{\dot \lambda}{\lambda}
+ O \left ( \frac{a^2H^2}{k^2} \right ) \right )
\dot \delta = 0.
\label{450}
\end{align}
Eq.~(\ref{450}) is quite different from that in quintessence model. There are
no oscillating solutions, moreover,
the quasi-static solutions are modified.
In fact, if we assume
\begin{equation}
\ddot \delta + 2H \dot \delta -\frac{\kappa ^2}{2} \rho \delta = 0,
\label{460}
\end{equation}
then \begin{align}
\ddddot \delta + \left ( 7H - \frac{\dot \lambda}{\lambda}  \right ) \dddot
\delta +
\left ( 16H^2 - 4 \kappa ^2 (\rho - 2 \lambda) -5H \frac{\dot \lambda}{\lambda}
\right ) \ddot \delta \nonumber \\
+ \frac{3}{2} \left ( 8H^3 -3 \kappa ^2 H (\rho - 4 \lambda) -4 H^2 \frac{\dot
\lambda}{\lambda} + \kappa ^2 \rho \frac{\dot \lambda}{\lambda}
   \right ) \dot \delta
   - \frac{\kappa ^4}{2} \rho \lambda \delta= 0.
\label{470}
\end{align}
Therefore, the solution which satisfies Eq.~(\ref{460}) is not the solution of
Eq.~(\ref{450}) except for the cases $\rho =0$ or $\lambda =0$ or $\delta =0$.
It is necessary to be careful because there is an ambiguity to express
Eq.~(\ref{470}) due to adding the terms proportional to
Eq.~(\ref{460}) or the derivatives of Eq.~(\ref{460}) are allowed. 
In $k$-essence model, the growth equation of the matter density perturbation 
is also represented by fourth order equation\cite{Bamba:2011ih} as in Eq.~(\ref{450}). 
However, in general, the sound speed read off from the field equation (\ref{PFE}) is not zero 
in the $k$-essence model, so there are no modifications to the quasi-static solution.
Thus, the modification to the quasi-static solution in Eq.~(\ref{450}) is the peculiar feature in this theory. 
However, in the matter dominant era or in $\lambda$ dominant era,
the term $ - \kappa ^4 \rho \lambda \delta /2$ is less than the other terms in
Eq.~(\ref{470}).
Therefore, we need to consider the era where there is almost same amount of
dark energy and dark matter
such as the current universe to distinguish Eq.~(\ref{450}) from
Eq.~(\ref{470}).
Whereas, in the case of $V_{,\phi}=0$, Eq.~(\ref{450}) recovers
the differentiated equation of (\ref{totsr}), correctly.

The procedure to obtain Eq.~(\ref{450}) is
\begin{enumerate}
\item differentiating Eq.~(\ref{440}) with respect to $t$
\item eliminating the terms proportional to $\delta \dddot \phi$
from the equation derived at the first step by using Eq.~(\ref{420})
\item differentiating  the equation derived at the second step with respect to
$t$
\item eliminating the terms proportional to $\delta \dddot \phi$
from the equation derived at the third step by using Eq.~(\ref{420})
\item differentiating the equation derived at the fourth step with respect to
$t$
\item eliminating the terms proportional to $\delta \dddot \phi$
from the equation derived at the fifth step by using Eq.~(\ref{420})
\item eliminating the terms proportional to $\delta \ddot \phi$
from the equations derived at the second, fourth, and sixth step by using
Eq.~(\ref{440})
\item eliminating the terms proportional to $\delta \dot \phi$ and $\delta
\phi$
by using the equations derived at the seventh step.
\end{enumerate}
\subsubsection{Mimetic wCDM}
\begin{figure}
\begin{center}
\includegraphics[clip, width=0.8\columnwidth]{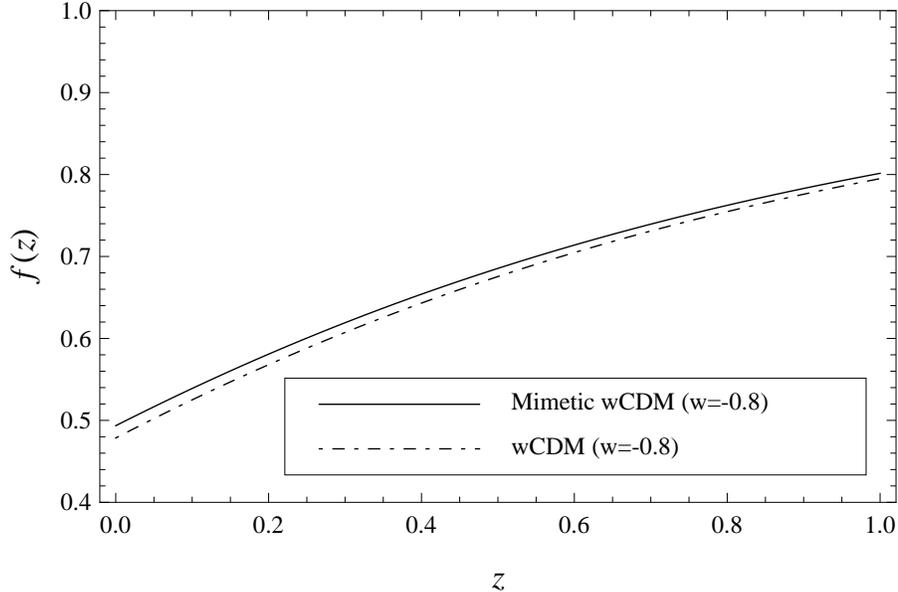}
\end{center}
\caption{Difference between mimetic wCDM and wCDM model in the growth rate
function $f(z)$ when $w=-0.8$. }
\label{f2}
\end{figure}

\begin{figure}
\begin{center}
\includegraphics[clip, width=0.8\columnwidth]{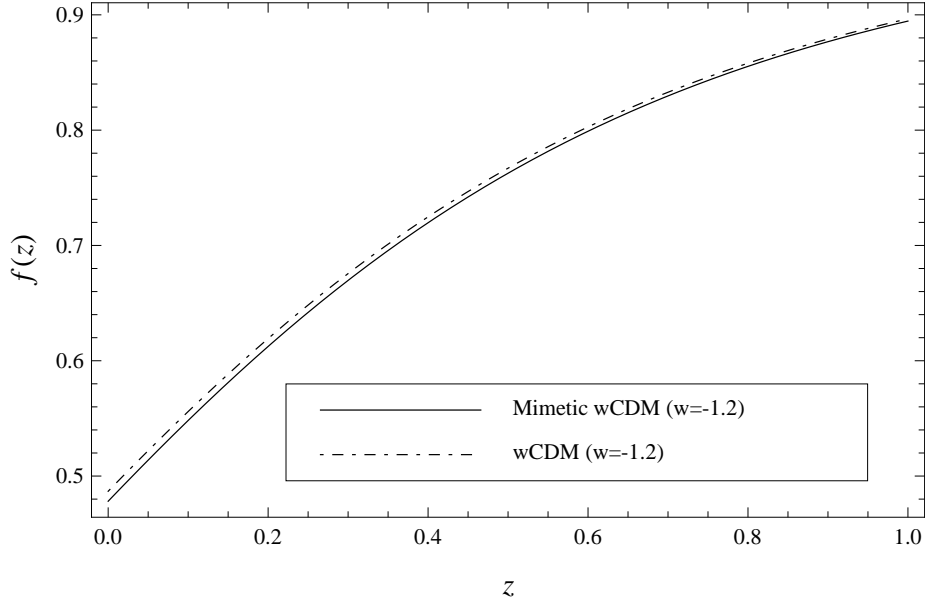}
\end{center}
\caption{Difference between mimetic wCDM and wCDM model in the growth rate
function $f(z)$ when $w=-1.2$. }
\label{f3}
\end{figure}

Let us evaluate the difference between Eqs.~(\ref{450}) and (\ref{460}) in the
case that mimetic matter
behaves like wCDM model. For a correct evaluation, we will not use
Eq.~(\ref{appV}) but use Eqs.~(\ref{wlambda}) and
(\ref{wpotential}).
If we utilize $N \equiv \ln a$ instead of time $t$, we can describe all the
functions in Eq.~(\ref{450}) as the functions of $a$ by
using Eqs.~(\ref{FL00}), (\ref{FLii}), (\ref{sol1}), (\ref{wlambda}), and
(\ref{wpotential}).
Therefore, Eq.~(\ref{450}) can be solved at least by numerical calculations.
\begin{figure}
\begin{center}
\includegraphics[clip, width=0.8\columnwidth]{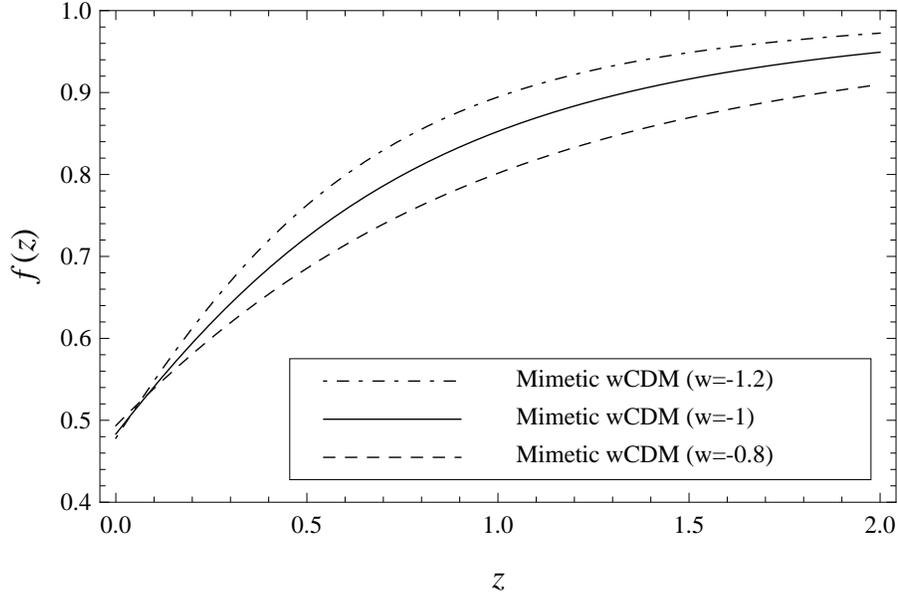}
\end{center}
\caption{Redshift dependence of the growth rate function $f(z)$ in mimetic
wCDM.
The dot-dash line, the solid line, and the dashed line express $f(z)$ in the
cases $w=-1.2$, $w=-1$, and $w=-0.8$, respectively.
$\Omega _{w0}=0.73$, $\Omega _{dm0}=0.23$, and $\Omega _{b0} =0.04$ are assumed
to depict the above figure.
Initial conditions are $f''(100)=f'(100)=0$ and $f(100)=1$. }
\label{f1}
\end{figure}
Fig.~\ref{f1} shows the changes of the growth rate function $f(z) \equiv d \ln
\delta (z)/ dN$, with
respect to $z$, where $z=-1+ 1/a= -1+ \mathrm{e}^{-N}$.
Figs.~\ref{f2} and \ref{f3} express the differences between mimetic wCDM and
wCDM model in the growth rate function.
Here, $\Omega _{w0}=0.73$, $\Omega _{dm0}=0.23$, and $\Omega _{b0} =0.04$,
where $\Omega _{X0} \equiv \kappa ^2 \rho _{X0}/(3H_0^2) $
is the current energy fraction of the matter X, are assumed to depict the
figures. 
As the initial conditions, $f''(100)=f'(100)=0$ and $f(100)=1$ are used, 
because, in the matter dominant era, i.e. $V \simeq const.$, we could use Eq.~(\ref{totsr}) 
and can obtain $f''(z)=f'(z)=0$ and $f(z)=1$. 
As noted in Sec.~\ref{mdm} and at the beginning of Sec.~\ref{mmp}, there is no
difference between mimetic model and wCDM model when $w=-1$.
However, in the case $w \neq -1$ $\&$ $w<0$, there is difference which come
from the term $- \kappa ^4 \rho \lambda \delta /2$ in Eq.~(\ref{470}).
One can see in Figs.~\ref{f2} and \ref{f3} that the differences are small but
surely exist.
The term $- \kappa ^4 \rho \lambda \delta /2$ is not relevant at the matter
dominant era,
so we can also find that the differences become small as redshift z becomes
large.

\subsubsection{Dynamical Om(z)}
\begin{figure}
\begin{center}
\includegraphics[clip, width=0.8\columnwidth]{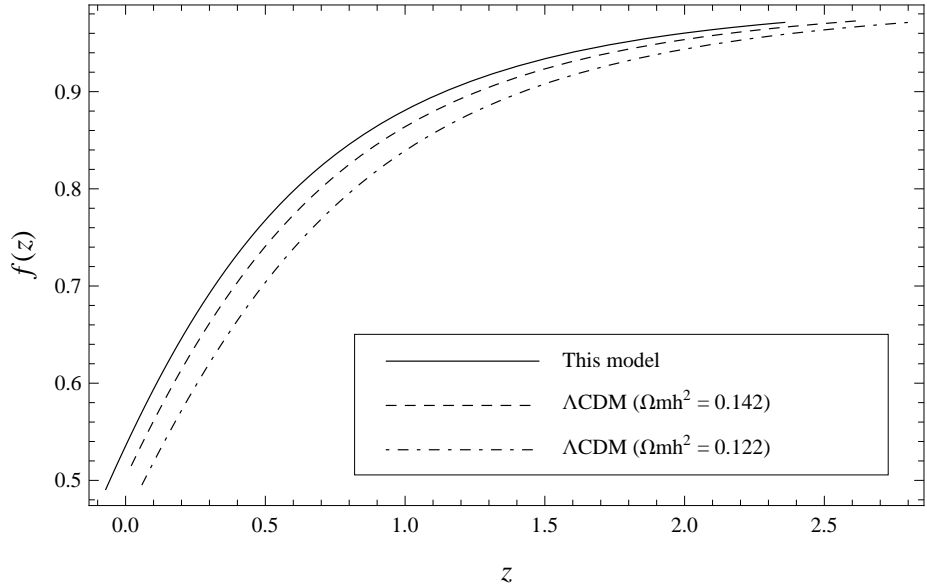}
\end{center}
\caption{The growth rate function $f(z)$ when the evolution of the background
space-time
is described by Eq.~(\ref{om4}). Solid, Dashed, and Dot-dashed curves represent
$f(z)$ in this model, $f(z)$ in the $\Lambda$CDM model with $\Omega _{m0}h^2 =
0.142$,
and $f(z)$ in the $\Lambda$CDM model with $\Omega _{m0}h^2 = 0.122$,
respectively. Initial conditions are $f''(100)=f'(100)=0$ and $f(100)=1$.}
\label{f4}
\end{figure}

The behavior of the growth rate function $f(z)$ when the background solution of
the Universe is
given by Eq.~(\ref{om4}) is expressed in Fig.~\ref{f4}.
$H_0=100h kms^{-1}Mpc^{-1}$, $h=0.7$ is assumed to depict Fig.~\ref{f4}.
Although $Om(z)$ is a smoothly changing function from $0.142/h^2$ to
$0.122/h^2$ when we
consider $z<1089$, however, the growth rate function $f(z)$ is larger than
that of the $\Lambda$CDM model with $\Omega _{m0}h^2 = 0.142$.
This effect comes from the difference between Eqs.~(\ref{450}) and (\ref{460}).
In fact, if we evaluate $f(z)$ for a constant $Om$ model with $Om h^2 = 0.142$
in Eq.~(\ref{450}),
we have a larger value than that of the dynamical $Om (z)$ (\ref{om4}).

\section{On the possibility of the phantom -- non-phantom transition
\label{phantom}}
\subsection{Reconstruction of the background evolution}
In the quintessence model, the value of EoS parameter $w=p _\phi /\rho _\phi$
is constrained as $-1<w<1$ so that
phantom behavior $w<-1$ cannot be realized.
Here, we  formally consider the possibility to construct a model, which can
describe the transition from
$-1<w$ to $w<-1$ or from $w<-1$ to $-1<w$.
If the Hubble rate $H(t)$ is given by the following function,
\begin{equation}
H(t) = \frac{\alpha}{3} (T_0+ t)^3 - \beta (T_0 + t) + \gamma, \; \gamma = -
\frac{\alpha}{3}T_0 ^3 + \beta T_0,
\label{UID}
\end{equation}
\begin{equation}
H(t) = h_0 + h_1 \sin (\nu t),
\end{equation}
\begin{equation}
H(t) = c_0 \left ( \frac{1}{t} + \frac{1}{t_s - t} \right ), \label{ex3}
\end{equation}
or
\begin{equation}
H(t) =t_2 \left ( \frac{1}{t_0^2 - t^2} + \frac{1}{t_1^2+t^2} \right ),
\end{equation}
then for the case, $0<t<t_-$ or $t_+ < t$, $h_1 \nu \cos (\nu t) >0$, $t_s/2 <
t < t_s$, or $t'_- < t <0$ or $t>t'_+$,
where $t_{\pm} \equiv -T_0 \pm \sqrt{\beta / \alpha}$ and $t'_{\pm} \equiv \pm
\sqrt{(t_0^2-t_1^2)/2}$,
effective EoS parameter $w_\mathrm{eff}=-2\dot H/(3H^2)-1$ satisfies
$w_\mathrm{eff}<-1$, respectively\cite{Nojiri:2005pu}.
Here, $T_0, h_1$, and $\nu$ are real valued
constants and $\alpha , \beta , h_0, c_0 , t_s, t_0, t_1 < t_0$ and $t_2$ are
positive constants.
Eq.~(\ref{UID}) could also describe the unified model of the inflation and dark
energy
because there are two regimes $0<t<t_-$ and $t_+ < t$ satisfying
$w_\mathrm{eff}<-1$, the era $0<t<t_-$ and
the era $t_+ < t$ correspond to the inflation and a dominance of dark energy,
respectively.
The forms of the potential which cause the above expansion rates of the
Universe are obtained by using Eq.~(\ref{rec}) as
\begin{align}
V(\phi) = \frac{1}{\kappa ^2} \bigg [ \frac{\alpha ^2}{3} (T_0 + \phi) ^6 -2
\alpha \beta (T_0+ \phi )^4 +2 \alpha \gamma (T_0 + \phi )^3 \nonumber \\
   +(2 \alpha + 3 \beta ^2 ) (T_0 + \phi )^2 -6 \beta \gamma (T_0 + \phi ) +3
\gamma ^2 -2 \beta \bigg ],
\end{align}
\begin{equation}
V(\phi) = \frac{1}{\kappa ^2} \left [ 3h_0^2 +6 h_0 h_1 \sin (\nu \phi) + 2 h_1
\nu \cos (\nu \phi) +3 h_1^2 \sin ^2 (\nu \phi) \right ],
\end{equation}
\begin{equation}
V(\phi) = \frac{1}{\kappa ^2} \left [
   \frac{3c_0^2-2c_0}{\phi ^2} + \frac{6c_0^2}{\phi (t_s-\phi)} + \frac{2c_0 +
3c_0^2}{(t_s - \phi)^2} \right ] , \label{ex3V}
\end{equation}
and
\begin{equation}
V(\phi) = \frac{1}{\kappa ^2} \left [
   \frac{4t_2 \phi+3t_2^2}{(t_0^2-\phi ^2)} + \frac{6 t_2^2}{(t_0^2 - \phi
^2)(t_1^2 + \phi ^2)} + \frac{3 t_2^2 -4 t_2 \phi}{(t_1^2 + \phi ^2)^2}
\right ] ,
\end{equation}
respectively. Here, EoS parameter of the matter is assumed to be $0$. As seen
in Eq.~(\ref{wlambda}),
the phantom crossings in these examples are realized by the transition from
$\lambda >0$ to $\lambda <0$
when there is no usual matter. Hence, in mimetic matter model with potential
the phantom-divide transitions maybe realized by proper choice of the scalar
potential.
\subsection{Growth of the matter density contrast in phantom crossing model}
Let us consider the evolution behavior of the matter density contrast in the
model with Eq.~(\ref{ex3V}) as an example.
In this model, the Universe will experience the singular evolution in $H(t)$ if
$t_s$ is larger than the current time.
It is known that there may occur  several finite-time future singularity
scenarios of the Universe as \cite{Starobinsky:1999yw,singular,reconstruction} which are
classified as \cite{singular}:
\begin{itemize}
\item Type I (''Big Rip'') : For $t \to t_s$, $a \to \infty$,
$\rho_\mathrm{eff} \to \infty$, and $\left|p_\mathrm{eff}\right| \to \infty$.
This also includes the case of $\rho_\mathrm{eff}$, $p_\mathrm{eff}$ being
finite at $t_s$.
\item Type II (''Sudden'') : For $t \to t_s$, $a \to a_s$,
$\rho_\mathrm{eff} \to \rho_s$, and $\left|p_\mathrm{eff}\right| \to \infty$.
\item Type III : For $t \to t_s$, $a \to a_s$,
$\rho_\mathrm{eff} \to \infty$, and $\left|p_\mathrm{eff}\right| \to \infty$.
\item Type IV : For $t \to t_s$, $a \to a_s$,
$\rho_\mathrm{eff} \to 0$, $\left|p_\mathrm{eff}\right| \to 0$, and higher
derivatives of $H$ diverge.
This also includes the case in which $p_\mathrm{eff}$ ($\rho_\mathrm{eff}$)
or both of $p_\mathrm{eff}$ and $\rho_\mathrm{eff}$
tend to some finite values, while higher derivatives of $H$ diverge.
\end{itemize}
Singularity in the case under consideration corresponds to Type I or Type III.
When $c_0=2/3$ and $t_s = 4 \times 10^{10}$ years, we have a solution of
Eq.~(\ref{ex3}) as
\begin{equation}
a(t) = c_1 \left ( \frac{t}{t_s-t} \right )^\frac{2}{3}, \label{ex3a}
\end{equation}
where $c_1$ is an integration constant. It is chosen   as $c_1=1.6$
to comply with the observed background evolution of the Universe. In this case,
$w_\mathrm{eff}$ is simply expressed as $w_\mathrm{eff}= -2t/t_s$.
The evolution behavior of the matter density perturbation is changed around the
singularity
because the approximation $\vert a^2H^2/k^2 \vert \ll 1$ no longer holds in
Eq.~(\ref{450}).
\begin{figure}
\begin{center}
\includegraphics[clip, width=0.8\columnwidth]{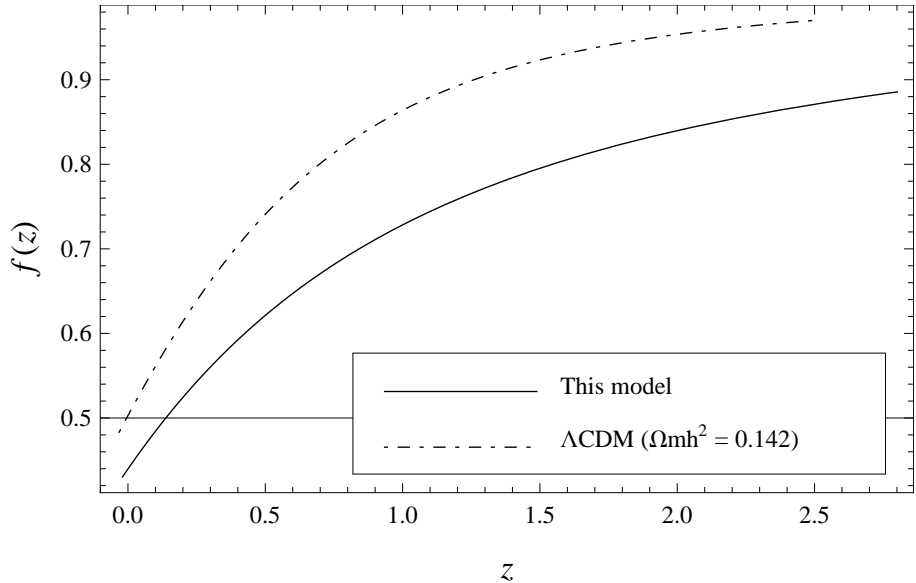}
\end{center}
\caption{The growth rate function $f(z)$ in the model (\ref{ex3V})}
\label{f5}
\end{figure}
\begin{figure}
\begin{center}
\includegraphics[clip, width=0.8\columnwidth]{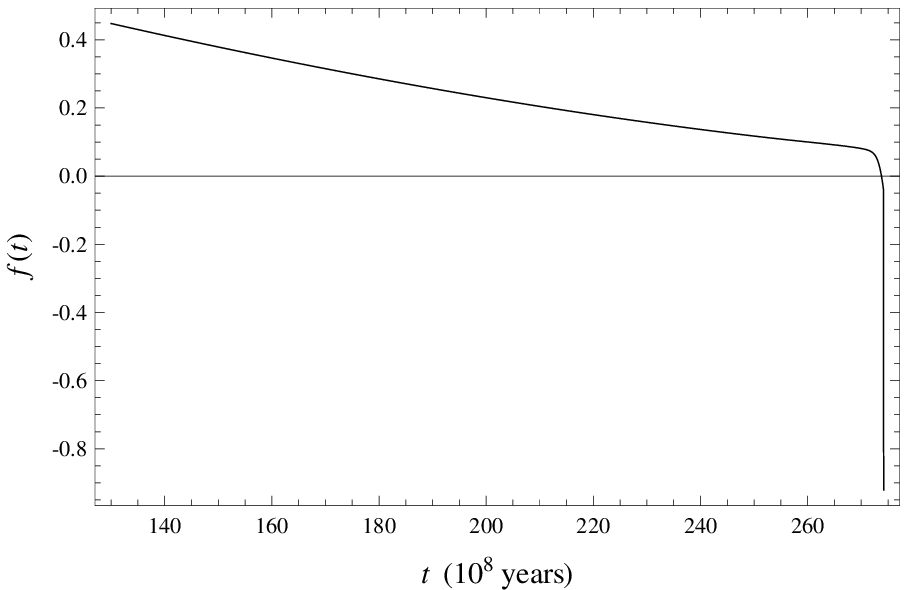}
\end{center}
\caption{The future evolution of the growth rate function $f(t)$ in the model
(\ref{ex3V})}
\label{f6}
\end{figure}
In Fig.~\ref{f5}, the behavior of the growth rate function $f(z)=d \ln \delta
(z) / d N (z)$ until the current time is expressed.
On the other hand, the future evolution of $f(t)$ is expressed in
Fig.~\ref{f6}.
Fig.~\ref{f6} shows singular behavior in $f(t)$ around $t=2.7 \times 10^{10}$
years,
although the singularity in $a(t)$ and $H(t)$ occurs at $t=4 \times 10^{10}$
years.
The reason why such a behavior of $f(t)$ occurs is instability of the
perturbations of phantom matter even at the classical level.
To be concrete, the flip of the sign of the kinetic term would cause the
instability.
A decay of perturbations much earlier of future singularity found in
\cite{Astashenok:2012iy} is just
the same as this instability 
(for discussion of problems of phantom-divide transitions in 
perturbations, see \cite{Vikman:2004dc,Hu:2004kh,Caldwell:2005ai}). 

\section{Conclusions}
In this paper, we have investigated the evolution of the matter density
perturbation in mimetic matter model.
In Sec.~II, the reconstruction, the way to construct a model which can describe
arbitrary evolution history of the Universe,
has been explicitly shown and some examples, including mimetic wCDM model and
evolving Om(z) model, have been given.
At the beginning of Sec.~III, the growth rate of the matter density
perturbation has been investigated when potential term is not dynamical.
As a result, it has been shown that mimetic dark matter also behaves as CDM at
the perturbation level.
In the case of dynamical potential, it has been shown that the differential
equation of the matter density perturbation
is changed.
In general, this equation is different from that in the $\Lambda$CDM model and
the quasi-static equation of the matter density perturbation
in quintessence model.
We have presented the behavior of the growth rate function in
mimetic wCDM model and in a model of dynamical $Om(z)$ as examples.
The difference seems not to happen in mimetic wCDM, but it clearly appears in
dynamical $Om(z)$.
In Sec.~IV, we have proved the possibility to realize phantom -- non-phantom
transition in mimetic matter model.
It has been shown that phantom -- non-phantom transition is possible because
the sign of kinetic term of the scalar field
is determined by the sign of the Lagrange multiplier.
However,  usually phantom brings an instability.
Such an instability occurs not only as quantum effect but also as classical
effect\cite{Dubovsky:2005xd,Woodard:2006nt,Sawicki:2012pz,Cline:2003gs,Garriga:2012pk,Elizalde:2004mq,Nojiri:2005sr}, 
and it breaks the perturbation theory as is explicitly demonstrated.

It would be interesting to study the accelerating universe evolution and
cosmological perturbations in the generalizations of mimetic matter model which
include mimetic $F(R)$ gravity \cite{no,sar}, higher-derivative mimetic
model\cite{mir} or its further generalizations\cite{mm}.
In this case, we have not only scalar potential but also extra function like
$F(R)$ or higher-derivative term. Hence, the cosmological reconstruction can be
executed quite easily. This will be studied elsewhere.

   \section*{Acknowledgments}
JM would like to acknowledge a suggestion from A.~A.~Starobinsky.
The work was
supported by the Russian Government Program of Competitive Growth of Kazan
Federal
University and, partially, by the Russian Foundation for Basic Research grants
No. 14-02-
00598.
The research by SDO has been supported in part by MINECO (Spain), projects
FIS2010-15640 and FIS2013-44881.

\end{document}